\documentclass[iop,revtex4]{emulateapj}
\usepackage{natbib,xspace,color}
\usepackage{verbatim}
\usepackage{ gensymb }
\usepackage{ mathrsfs }
\usepackage{amsmath}
\usepackage{amssymb }
\newcommand{\kms}{km~s$^{-1}$}

\newcommand{\ang}{\mbox{\AA} }
\newcommand{\mg}{$M_{\rm{g}}$}
\newcommand{\dgf}{$f_{\rm{DG}}$}
\newcommand{\mpcsq}{M$_{\odot}$pc$^{-2}$}

\newcommand{\msun}{M$_{\odot}$}
\newcommand{\cmsq}{cm$^{-2}$}
\newcommand{\cc}{cm$^{-3}$}

\newcommand{\sigsfr}{$\Sigma_{SFR}$}
\newcommand{\sigmg}{$\Sigma_{\rm{g}}$}
\newcommand{\co}{$^{12}$CO}
\newcommand{\coa}{$^{13}$CO}

\newcommand{\vlsr}{$V_{\rm{LSR}}$}
\newcommand{\dkin}{$D_{\rm{kin}}$}
\newcommand{\htwo}{\mbox{${\rm H}_2$}}

\newcommand{\sfr}{$SFR$}

\shorttitle{The Dense Gas Mass Fraction in Molecular Clouds}
\shortauthors{Battisti \& Heyer}

\begin{document}

\title{The Dense Gas Mass Fraction of Molecular Clouds in the Milky Way}
\author{A. J. Battisti\altaffilmark{1}, 
M. H. Heyer\altaffilmark{1}
}


\altaffiltext{1}{Department of Astronomy, University of Massachusetts, Amherst, MA 01003, USA; abattist@astro.umass.edu}

\begin{abstract}
The mass fraction of dense gas within giant molecular clouds (GMCs) of the Milky Way is investigated using \coa\ data from the FCRAO Galactic Plane Surveys and the Bolocam Galactic Plane Survey (BGPS) of 1.1~mm dust continuum emission.  A sample of 860 compact dust sources are selected from the BGPS catalog and kinematically linked to 344 clouds of extended ($>$3\arcmin) \coa\ J=1-0 emission.  Gas masses are tabulated for the full dust source and subregions within the dust sources with mass surface densities greater than 200 \mpcsq, which are assumed to be regions of enhanced volume density. Masses of the parent GMCs are calculated assuming optically thin \coa\ J=1-0 emission and LTE conditions. The mean fractional mass of dust sources to host GMC mass is $0.11_{-0.06}^{+0.12}$. The high column density subregions comprise $0.07_{-0.05}^{+0.13}$ of the mass of the cloud. Owing to our assumptions, these values are upper limits to the true mass fractions. The fractional mass of dense gas is independent of GMC mass and gas surface density. The low dense gas mass fraction suggests that the formation of dense structures within GMCs is the primary bottleneck for star formation. The distribution of velocity differences between the dense gas and the low density material along the line of sight is also examined.  We find a strong, centrally peaked distribution centered on zero velocity displacement. This distribution of velocity differences is modeled with radially converging flows towards the dense gas position that are randomly oriented with respect to the observed line of sight. These models constrain the infall velocities to be 2-4 \kms\ for various flow configurations.
\end{abstract}
\keywords{ISM: clouds --- ISM: kinematics and dynamics --- stars: formation}

\section{Introduction}
The production of newborn stars in galaxies is regulated by a sequence of interstellar processes that span a range of spatial scales, environmental conditions, and gas phases. The development of molecular clouds from the diffuse, atomic, interstellar medium is impacted by spiral density waves, mid-plane pressures owing to the overlying weight of stars and gas, the interstellar ultraviolet radiation field, magneto-turbulence, and expanding shells from the cumulative effects of supernovae and HII regions.  Jeans' and thermal instabilities, ambipolar diffusion, and shocks from converging, super-Alfvenic flows are drivers of the fragmentation of molecular clouds into dense ($\sim10^3$-$10^4$ \cc) filaments and clumps.  From these filament structures, even higher density ($>10^4$ \cc), compact structures emerge from the actions of gravity, ambipolar diffusion, instabilities, and turbulent shocks. Such high volume density configurations are the basic unit from which newborn stars form. Given the large range of scales and physical processes, a composite view of star formation is required, which includes information from nearby galaxies and well-resolved star forming regions in the Milky Way \citep{kennicutt12}.  

Star formation in galaxies is frequently parameterized by the scaling relationship between the rate of star formation, \sfr, and the mass of available neutral, interstellar gas, \mg.  More typically, the relationship is cast between the surface density of the star formation rate, \sigsfr, and gas surface density, \sigmg\ \citep{kennicutt89}.  The supplies of atomic and molecular gas mass are respectively traced by HI~21~cm line emission and the low-J rotational lines of CO. The relationship between \sigsfr\ and \sigmg\ is well described by a power law with an index, $\gamma=\log(\Sigma_{SFR})/\log(\Sigma_g)$ ranging from sub-linear \citep{shetty13}, to linear \citep{bigiel08,leroy13} to super-linear \citep{kennicutt98}.  Comparable indices are found if one only considers the molecular gas surface density as derived from CO, which primarily tracks the low density substrate of molecular clouds \citep{wong2002, heyer04, bigiel08}.  Followup studies that examine the surface density of dense gas using HCN $J=1-0$ emission or the higher rotational transitions of CO identify a linear correlation with \sigsfr\ \citep{gao04}, which emphasizes that star formation is regulated by the development of dense gas configurations within molecular clouds.

In recent years, there have been several efforts to connect the star formation properties of resolved, Galactic star forming regions to these extragalactic scaling relationships.  \citet{wu05} extended the analysis of \citet{gao04} to Galactic star forming regions and found a similar slope and amplitude between the far infrared and HCN $J=1-0$ luminosities as found in galaxies.  Infrared data from the Spitzer Space Telescope enabled direct counting of young stellar objects emerging from nearby molecular clouds.  \citet{heiderman10} examined the numbers of Class 1 and Flat SED young stellar objects (YSOs), lying within varying ranges of gas surface density, as traced by infrared-derived  extinctions for 20 star forming regions located within 0.5 kpc of the Sun.  They found a linear correlation between \sigsfr\ and \sigmg\ when \sigmg\ $>$ 130 $M_\odot \rm{pc}^{-2}$ but with an amplitude significantly larger than the value determined for galaxies. \citet{gutermuth11} similarly tabulated the numbers of YSOs (Class 1 and Class 2) identified within 8 nearby ($<$1 kpc) star forming regions. Using infrared-derived extinction, as a measure of \sigmg, they derive \sigsfr $\propto$ $\Sigma_g^2$ and a large offset of \sigsfr\ for a given surface density, \sigmg.  \citet{lada10,lada12} derive a linear  relationship between \sfr\ and molecular mass for a set of local clouds with the same mass fraction of dense gas, \dgf. They propose \sigsfr$ \propto $\dgf\sigmg, as the fundamental relationship governing star formation. 

In this paper, we investigate the dense gas mass fraction, \dgf, for a large number of GMCs that reside in the inner Galaxy where most star formation takes place in the Milky Way. Specifically, we address the range of the dense gas mass fraction and the degree to which \dgf\ varies with GMC mass and surface density.  Our study utilizes extensive surveys of the Galactic Plane in \coa\ and 1.1~mm dust emission, along with published spectral line observations (HCO$^+$ $J=3-2$, N$_2$H$^+$ $J=3-2$, or NH$_3$ (1,1)) towards the 1.1~mm dust continuum sources that probe dense gas \citep{dunham11b, schlingman11}.  Each survey samples a key phase and sequence of the star formation process. \coa\ emission provides the distribution and kinematics for the low to moderate density gas ($n \sim 10^3$~\cc) substrate of the cloud. The 1.1~mm dust emission traces regions of enhanced surface density and under restrictive conditions, high volume density gas ($n\gtrsim10^{3.5}$~\cc), corresponding to sites of massive clumps and filaments within these clouds.  In \S2, the observational data sets, source selection, and analysis methods are described.  In \S3, we summarize the cloud and dust source properties. The dense gas mass fraction is discussed in \S4. In \S5, we examine and model the motions of low density gas circumscribing the dust continuum source in the context of converging flows. 

\section{Data and Analysis}
\subsection{\coa\ $J=1-0$ Data}
The Boston University-Five College Radio Astronomy Observatory (BU-FCRAO) Galactic Ring Survey \citep[GRS;][]{jackson06} and the Exeter-FCRAO Survey of the Galactic Plane (unpublished) imaged the \coa\ $J=1-0$ emission along the Galactic Plane with the 14~meter telescope of the Five College Radio Astronomy Observatory (FCRAO).  The GRS covers the Galactic longitude range $l=18^\circ-55.7^\circ$ and latitude range of $|b|\le1^\circ$ with a median sensitivity of 0.28~K in main beam temperature units within 0.2~\kms\ wide channels.  The Exeter-FCRAO Survey of \co\ and \coa\ $J=1-0$ emission spans longitudes 55$^\circ$ to 100$^\circ$ and 140$^\circ$ to 19$5^\circ$ with varying latitude coverage but no less than $\pm$1$^\circ$.  The median sensitivity of the \coa\ data within 0.13~\kms\ wide channels is 0.5~K in main beam temperature units.  Both surveys are fully sampled, so the angular resolution of the \coa\ data used in this study is 47\arcsec\ corresponding to the full width at half-maximum (FWHM) of the 14~m telescope at 110.201~GHz.

\subsection{1.1 mm Dust Continuum Emission}
The Bolocam Galactic Plane Survey (BGPS) imaged the 1.1 mm dust continuum emission between Galactic longitudes of $-10^\circ < l < 90^\circ$ and latitudes $|b|\le0.5^\circ$ with the Caltech Submillimeter Observatory (CSO) \citep{aguirre11}.  The 1-$\sigma$ surface brightness sensitivity ranges from 11-53 mJy/beam and the FWHM angular resolution is 33\arcsec.  The source catalog derived from the BGPS is described by \citet{rosolowsky10}.  For point sources, the catalog is 99\% complete for fluxes greater than 0.2-0.4 Jy over the range of Galactic longitudes used in this study.  As recommended by \citet{aguirre11}, all source fluxes listed in the catalog are multiplied by a factor of 1.5 to be consistent with source fluxes measured with other millimeter continuum instruments. 

\subsection{BGPS Source Selection}
We analyze a subset of BGPS sources from the catalog of \citet{rosolowsky10} with the following selection criteria.  Only sources with integrated 1.1~mm flux densities in excess of 0.75 Jy in the BGPS catalog, after the application of the 1.5 flux correction factor, are analyzed.  This limit is well above the flux level at which the catalog is 99\% complete \citep{rosolowsky10}.  The flux limited sample of BGPS sources imposes a bias to the most massive systems at a given distance.  For a dust temperature of 14~K, the minimum masses detectable at this flux level are 68, 273, 1095, and 2465 \msun\ at distances 2, 4, 8, and 12~kpc respectively. The BGPS dust continuum sources must overlap with areas and velocities covered by the FCRAO surveys. Therefore, sources with $l < 18.5^\circ$ or \vlsr $<$ -10~\kms\ are excluded.  These provisions limit our study to the population of molecular clouds residing within the disk of the Milky Way and Galactocentric radii, $2.8< R_{\rm{gal}} < 8.3$~kpc.  At least one spectroscopic detection of a dense gas tracer and corresponding velocity measurement from the studies of \citet{schlingman11} ($J=3-2$ of HCO$^+$, N$_2$H$^+$) or \citet{dunham11b} (NH$_3$ (1,1)) are required.  The velocity of dense gas selects which \coa\ velocity component along the line of sight (if more than one) is linked to the 1.1mm source and facilitates the isolation of the parent molecular cloud from the extended \coa\ emission.  In addition, these detections confirm the presence of high density ($n\ge 10^{3.5}$~\cc) gas that is necessary to excite these transitions. Such densities are comparable to values inferred by \citet{lada12} from zones with K-band extinctions in excess of 0.8 magnitudes in a sample of nearby star forming regions.  While the detection of these tracers indicates the presence of dense gas, the mean gas density may be lower if material is inhomogenously distributed.  

\subsection{Isolating GMCs with BGPS sources}
The GMCs linked to each selected BGPS source are identified by regions of  \coa\ $J=1-0$ emission.  In the Galactic plane, the isolation of CO emission from a singular cloud is frequently complicated by the blending of signal from nearby or unrelated clouds at or near the same velocity.  Such crowding is particularly severe near the maximum velocity for a given line of sight through the inner Galaxy.  A cloud identification algorithm is necessary to uniformly isolate the CO emission from the target cloud linked to the BGPS source.

Given the prior information of the dust source position and velocity, we have decomposed the \coa\ $J=1-0$ emission centered on each BGPS source in our sample.  The algorithm \texttt{CPROPS} \citep{rosolowsky06} is applied to an extracted \coa\ data cube with an area typically $0.5\times0.5$~deg$^2$ in angular extent centered on the position of the BGPS source and $\pm$20~\kms\ of the velocity of the dense gas tracer. The temperature threshold, $T_{\rm{thr}}$, is varied for each cloud to best isolate its emission from other nearby clouds or clouds at a similar velocity.  Since our intent is to define the global properties of the cloud based on \coa\ $J=1-0$ emission and not the hierarchical structure of the cloud, we do not decompose the cloud further into constituent clumps or fragments.

A three-dimensional ($l$,~$b$,~\vlsr) mask is generated from the set of voxels that define the cloud at a given threshold.  Based on this mask, \texttt{CPROPS} calculates the positional and velocity moments and integrated \coa\ emission while also accounting for the finite resolutions of the antenna beam and spectrometer.  While the cloud mask is defined at a given threshold of antenna temperature, the cloud properties are extrapolated to a common threshold of $T_{\rm{thr}}=0$~K.  Since there is subjectivity in the precise threshold that defines the cloud, which leads to uncertainties in the properties, we identify each cloud over a sequence of intensity thresholds that reasonably isolates the cloud.  The tabulated cloud properties and errors are assigned to the average and standard deviation of values calculated for the threshold sequence. The GMC properties and errors are listed in Table~1. 

The BGPS source that links the \coa\ cloud as a region of interest, is labeled the primary source for the cloud.  Since we are interested in all sources of dense gas in the cloud, any BGPS source from the catalog with a voxel position within the cloud mask is linked to that cloud.  In some cases, a secondary sources may not have a flux greater than the flux limit in the initial selection requirement but are included in the mass estimate of dense gas.  BGPS sources without velocity measurements but with fluxes $>$ 0.75~Jy and coincident with a local \coa\ intensity peak, are also linked to the cloud. 
 
\subsection{Resolving the Kinematic Distance Ambiguity}
In order to transform observed properties of clouds and BGPS sources (column density, angular size) into physical properties (mass, physical size), it is necessary to determine the distance to each cloud.  In this study, we use kinematic distances, \dkin, derived from the rotation curve of \citet{clemens85}. For locations in the inner Galaxy, there are two kinematic distances equally spaced on either side of the tangent point which have the same radial velocity along any line of sight. The ambiguity is resolved using the presence or absence of HI self-absorption and the size-velocity dispersion relation for molecular clouds. 

\citet{jackson02} demonstrated the utility of the presence (absence) of self-absorption of HI 21~cm line profiles to discriminate near (far) side clouds.  All molecular clouds retain a residual amount of cold ($T=30$~K) atomic hydrogen gas owing to the equilibrium between \htwo\ formation and destruction by cosmic rays \citep{goldsmith07}.  In addition, molecular clouds are circumscribed by a layer of atomic gas, that may be cold ($T\sim100$~K) or warm ($T\sim8000$~K) \citep{van89,visser09}.  The cold, neutral atomic gas within a molecular cloud or circumscribing the cloud absorbs 21~cm line emission from the near ubiquitous, warm, neutral atomic component of the interstellar medium (ISM) over the narrow velocity range of the cloud.  Accordingly, for a near-side molecular cloud, one would expect to observe self-absorption in the 21~cm line as there is a significant background of warm neutral gas at the velocity of the cloud near the far side location.  However, for a far-side cloud, there is warm neutral gas at the near side location  between the cloud and the observer.  So any self-absorption from the far side cloud itself is ``filled-in'' by warm, neutral gas at the near side velocity.  Therefore, for far-side clouds, the expectation is the absence of a self-absorption feature at the velocity of the molecular cloud.  

For each cloud identified by \coa, the 21~cm line emission from the Very Large Array (VLA) Galactic Plane Survey \citep[VGPS;][]{stil06} is examined for the presence or absence of self-absorption.  An average, 21~cm line ``ON'' profile is calculated from all spectra that lie within the \coa-defined mask of the cloud.  A corresponding, mean ``OFF'' profile is derived from all spectra circumscribing the \coa\ mask with the same size area as the defined cloud.  The presence of self-absorption at the velocity of the molecular cloud is visually assessed independently by the authors in this study for consistency and reliability.  

A secondary discriminator to resolve the distance ambiguity relies on the size-velocity dispersion relationship for molecular clouds. The cloud-to-cloud relationship between cloud size, $R$, and velocity dispersion, $\sigma_v$, results from the co-linear, or universal, velocity structure functions of clouds such that the quantity $v_\circ=\sigma_v/(\Sigma_{\rm{GMC}}R)^{1/2}$ is nearly constant for clouds residing within the disk of the Milky Way  \citep{larson81,brunt03,heyer&brunt04,heyer09}. The near-invariance of $v_\circ$ enables an estimate of the distance to the cloud, $D_{\rm{svd}}$, based on the measured velocity dispersion, surface density, and the angular radius of the cloud, $\theta$, in radians, 
\begin{equation}
D_{\rm{svd}} = \frac{1}{\theta} (\Sigma_{\rm{GMC}}/100 M_\odot pc^{-2})^{-1}  (\sigma_v/v_\circ)^2 \ \rm{pc} \ .
\end{equation}
where $v_\circ=0.7\pm0.07$ \kms~pc$^{-1/2}$ \citep{heyer09}.  While the fractional population variance precludes its use as an absolute measure of distance, it is sufficient to discriminate a near-side cloud from a far-side cloud for velocities well displaced from the terminal velocity for most longitudes.  For each cloud harboring a BGPS source, the \coa\ velocity dispersion and cloud size, evaluated at both the near and far side distances, are examined in relation to the constant value of $v_\circ$.  The cloud distance is assigned to the kinematic distance, \dkin, that most closely agrees with the constant, $v_\circ$.
 
Both the HI self-absorption and size-velocity dispersion methods are applied to resolve the near/far side distance ambiguity.  However, a higher weight is assigned to the HI self-absorption measure to establish a distance.  Sources for which the measured cloud velocity is within 10~\kms\ of the terminal velocity of the rotation curve for the source longitude, are assigned a distance corresponding to the tangent point position. We have compared our distance assignments to those derived by \citet{ellsworth2013} and \citet{roman-duval2010}. For common sources, our near and far side assignments agree with 83\% and 88\% of the assignments made by \citet{ellsworth2013} and \citet{roman-duval2010} respectively.  While these methods to resolve the near/far side ambiguity are not infallible, the consistency between three independent groups provides confidence that these assignments are not the limiting error in our results.

\subsection{Mass of \coa\ Clouds}
To calculate the GMC mass from \coa\ emission, we assume optically thin emission, a constant excitation temperature of 10~K, and local thermodynamic equilibrium (LTE) conditions  to account for population of molecules in the higher excitation states.  This provides a direct conversion from the \coa\ luminosity, $L_{13}$, in units of K~kms$^{-1}$~pc$^2$, to GMC mass 
\begin{equation}
M_{\rm{GMC}}=0.41 \left[ 6.2R_{\rm{gal}}+18.7\right] L_{13} \;\;\; M_\odot
\end{equation}
where the term in square brackets accounts for the varying $^{12}$C to $^{13}$C
abundance ratio with Galactocentric radius,  $R_{\rm{gal}}$  \citep{milam05}.  A constant \htwo\ to \co\ abundance of 2$\times$10$^4$ is assumed \citep{blake1987}  and a mean molecular weight of 1.36 is applied to account for the contribution from helium. The GMC masses and estimated random errors are shown in Table~1.

 Our assumptions of a constant excitation temperature, constant abundance, and optically thin \coa\ emission are clearly a simplification of the complex thermal and density structure of molecular clouds along the line of site that can lead to systematic errors. \citet{padoan00} show that deviations from LTE conditions can cause underestimations of the \coa\ true column densities by a factor ranging from 1.3 to 7. An excitation temperature of 10~K provides a reasonable reference value.  It is motivated by the distributions of \co\ excitation temperatures which peaks at a value of 7~K \citep{heyer09,roman-duval2010}.  For densities less than the critical density of the $J=1-0$ transition ($n_{crit}\sim 2000$~\cc), \coa\ is likely subthermally excited.  In this regime, which may contain a significant fraction of the cloud mass, the column densities derived using the thermalized excitation temperature underestimate the true column density \citep{dickman78}.  For example, if the true excitation temperature is 5~K, then the calculated column density assuming $T_{ex}=10$~K is $\sim50\%$ the value using the correct measure of excitation. Similarly, our assumption of optically thin emission for all lines of sight leads to an underestimation of the true column density by the factor $\int dv\ \tau(v)/\int dv\ (1-exp(-\tau(v))$. \citet{roman-duval2010} find the mean opacity integrated over the position-position-velocity volume of GMCs is 1.5. This mean opacity corresponds to a column density 1.9 times larger than the column density assuming $\tau=0$. The application of a constant abundance for all clouds and all lines of sight within clouds introduces additional errors. The assumed relative \htwo\ to \coa\ abundances are only valid in the strongly self-shielded interior of the cloud. A recent study by \citet{ripple13} shows that the \coa\ self-shielded regime accounts for only 50\% of the total \htwo\ mass of the Orion molecular cloud. Given our assumptions and the described effects, the derived values for $M_{\rm{GMC}}$ should be considered lower limits. Finally, the use of kinematic distances can also impact the reliability of cloud mass measurements. Non-circular motions induced by a spiral arm potential or random cloud motions can also lead to uncertainties of order 50\% in the mass estimates but these are expected to be random.

 A more complex error to quantify is the identification of the GMC at a given threshold. The error assigned to $M_{\rm{GMC}}$ and listed in Table~1 is the standard deviation of mass values determined from the sequence of antenna temperature thresholds for which the cloud is well distinguished from  extended signal. If the number of thresholds is less than 3, then the assigned error is the statistical error calculated by \texttt{CPROPS} through the bootstrap method \citep{rosolowsky06}.  

The surface density of the GMC is calculated from the expression, 
\begin{equation}
\Sigma_{\rm{GMC}}= \frac{M_{\rm{GMC}}} {\pi R_{\rm{GMC}}^2}
\end{equation}
and the mean volume density is estimated by 
\begin{equation}
<n_{\rm{GMC}}>= \frac{3M_{\rm{GMC}}}{4\pi \mu m_{\rm{H}_2} R_{\rm{GMC}}^3}
\end{equation}
where $R_{\rm{GMC}}$ is the effective radius of the cloud.

\begin{figure*}[htb]
\includegraphics[width=6.5in]{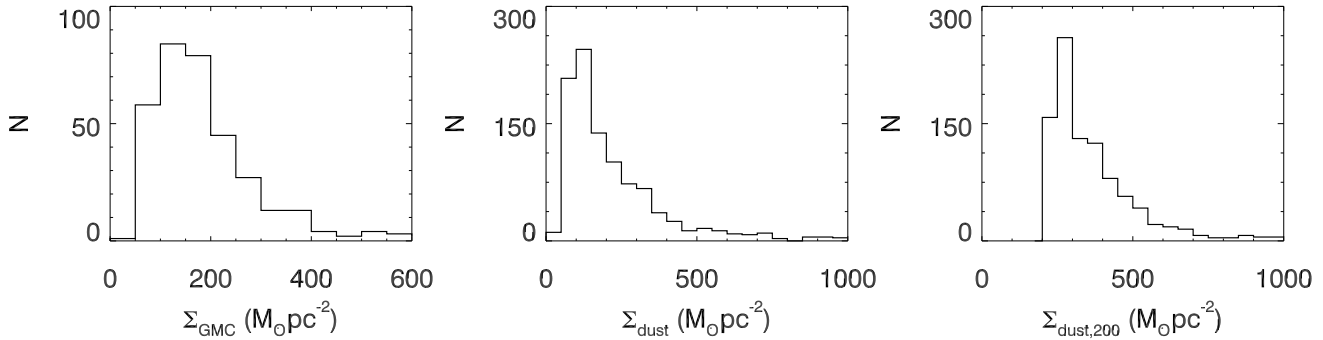}
\caption{The distributions of GMC surface density, $\Sigma_{\rm{GMC}}$, 
dust source surface density, $\Sigma_{\rm{dust}}$, and $\Sigma_{\rm{dust,200}}$ 
derived from source pixels with surface densities greater than 200 \mpcsq.}
\label{fig1}
\end{figure*}
\begin{figure*}[htb]
\includegraphics[width=6.5in]{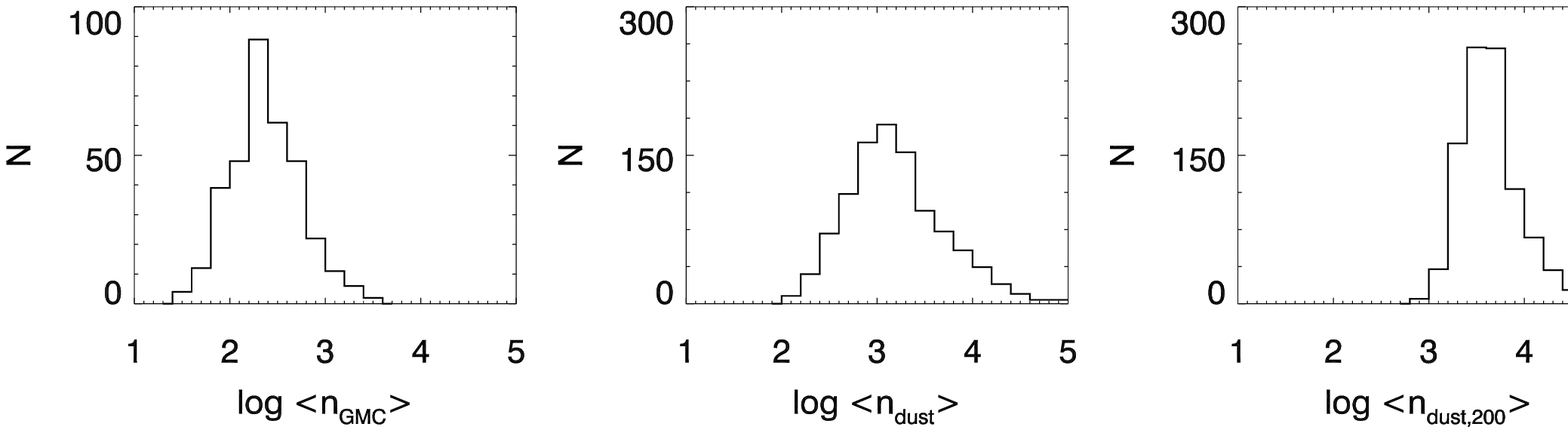}
\caption{The distributions of GMC mean volume density, $<n_{\rm{GMC}}>$, 
dust source volume density, $<n_{\rm{dust}}>$, and $<n_{\rm{dust,200}}>$.}
\label{fig2}
\end{figure*}

\subsection{Mass of BGPS Sources}
The mass of each BGPS source, $M_{\rm{dust}}$, can be estimated from its integrated flux density, $S_{1.1\rm{mm}}$, assuming that the dust emission at 1.1~mm is optically thin and well characterized by a single temperature and opacity,
\begin{equation}
M_{\mathrm{dust}}= \frac{D_{\rm{kin}}^2 S_{1.1\rm{mm}}}{B_{1.1\rm{mm}}(T_{\rm{dust}})\kappa_{1.1\rm{mm}}},
\end{equation}
where \dkin\ is the distance to the cloud harboring the BGPS source, $B_{1.1\rm{mm}}$ is the Planck function evaluated at $\lambda=1.1$~mm and dust temperature, $T_{\rm{dust}}$, and $\kappa_{1.1\rm{mm}}=0.0114$~cm$^2$~g$^{-1}$ is the dust opacity per gram of gas \citep{ossenkopf94} and includes a gas-to-dust ratio of 100. The dust column density, mass and respective uncertainties are determined from a Monte-Carlo simulation of equation~5 that samples a temperature distribution and source flux density distribution based on the measured flux density and flux density error listed in the BGPS catalog and assumed to follow a Gaussian function. All dust masses and estimated random errors are listed in Table~1. We assume a Gaussian distribution of dust temperatures centered on $T_{\rm{dust}}=14$~K with a dispersion of 4~K, which approximates the variation of dust temperatures found in BGPS sources by \citep{eden13}. We acknowledge a single dust temperature oversimplifies the distribution of thermal energy within a dust source. Temperature gradients along the line of sight arise from external and internal heating sources that can bias the estimate of a mean dust temperature along the line of sight to higher values and lead to an underestimation of the dust column density \citep{shetty09}.  For the BGPS sources studied here, an embedded massive protostar or young stellar cluster can raise the local dust temperature to 40-100 K within a limited volume of the dust source.  By assuming a lower dust temperature throughout the source, the derived mass from the dust continuum emission overestimates the true mass.   Also, we do not include uncertainties or a finite range of values in $\kappa_{1.1\rm{mm}}$ when calculating dust masses.  \citet{ossenkopf94} estimate that $\kappa_{1.1\rm{mm}}$ should not vary by more than a factor of two, which would similarly impact the dust mass error.

The mass surface density and mean volume density are
\begin{equation}
\Sigma_{\rm{dust}}= \frac{M_{\rm{dust}}} {\pi R_{\rm{dust}}^2}
\end{equation}
and 
\begin{equation}
<n_{\rm{dust}}>= \frac{3M_{\rm{dust}}} {4\pi \mu m_{\rm{H}_2} R_{\rm{dust}}^3}
\end{equation}
where $R_{\rm{dust}}$ is the effective radius of the dust continuum source.  For those sources that are unresolved along one or both of the source axes, the radius is assigned an upper limit corresponding to angular size of 14\arcsec\ at the distance of the source.

\begin{figure*}[htb]
\includegraphics[width=6.5in]{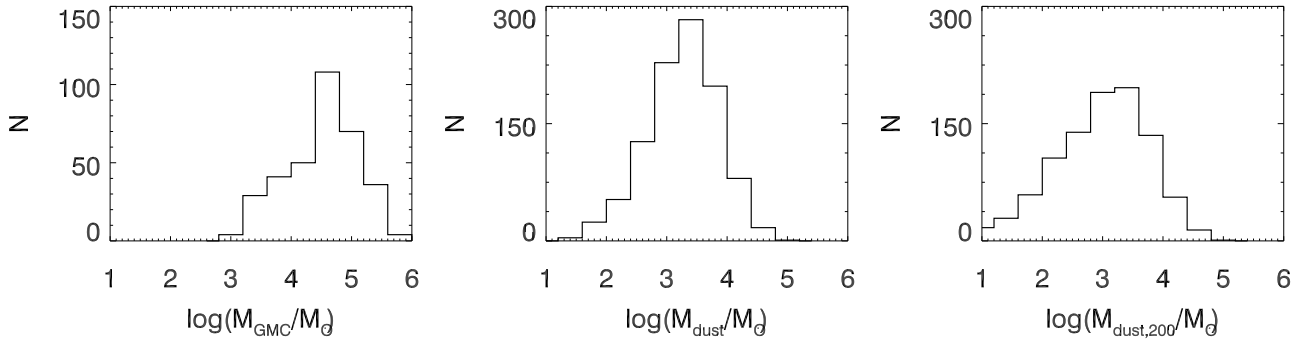}
\caption{The distributions of GMC mass, $M_{\rm{GMC}}$, the dust masses (primary and secondary) in the clouds, $M_{\rm{dust}}$, and the mass residing in high column density lines of sight $M_{\rm{dust,200}}$.  
}
\label{fig3}
\end{figure*}

\section{Results}
\subsection{BGPS-GMC Associations}
The dust continuum source requirements described in \S2.3 selects 437 primary sources from the BGPS catalog. An additional 578 secondary BGPS sources whose $l,b$ coordinates reside within the projected mask of these isolated GMCs are included in the accounting of the total dust mass. Of the 437 GMCs identified, 212 contain a singular dust source, 100 contain 2 dust sources, 52 contain 3 sources, and 73 contain 4 or more sources. To ensure a sample of well-defined clouds, we consider only those GMCs for which $M_{\rm{GMC}}/\sigma(M_{\rm{GMC}}) > 2$, where $\sigma(M_{\rm{GMC}})$ is the standard deviation of GMC mass values calculated over the sequence of antenna temperature thresholds that define the cloud. This condition is imposed to ensure that the identification of the GMC has converged. 344 objects satisfy this requirement. For this sub-sample of GMCs, 152 contain a singular dust source, 81 contain 2 dust sources, 43 contain 3 sources, and 68 contain 4 or more sources.

The nature of the selected BGPS sources and isolated clouds is revealed, in part, by the distributions of mass surface densities, mean volume densities, and masses.  Figure~\ref{fig1} shows the distributions of mass surface density for the GMCs, and all dust sources, $\Sigma_{\rm{dust}}$. The sharp lower cutoff for $\Sigma_{\rm{GMC}}$ is a result of the surface brightness threshold used to define the cloud, the lowest of which is 3 times the channel rms of the data. The mass surface densities for this sample of clouds are comparable to those determined by \citet{roman-duval2010} but higher than the values derived by \citet{heyer09}, who used the more extended cloud boundaries defined by \citet{solomon1987} based on \co\ emission.  In these cases, \coa\ emission does not fill the area resulting in lower surface densities. Surprisingly, the dust surface densities 
are comparable to the corresponding cloud surface densities.  For GMCs with small angular sizes, this equivalence is due to the median filtering of the Bolocam pipeline to remove atmospheric signal (discussed in more detail in \S4) that simply recovers the cloud mass.  For other dust sources the condition $\Sigma_{\rm{dust}}\sim\Sigma_{\rm{GMC}}$ may reflect a clumpy distribution of dense sub-fragments.  The clumpy nature of the dust sources is supported by the derived mean volume densities shown in Figure~\ref{fig2} and those tabulated by \citet{dunham11b} and \citet{schlingman11}.  The volume densities of the dust sources are larger than those of the GMCs by factors of 3-5 yet smaller than the excitation requirements of HCO$^+$ $J=3-2$ line and the NH$_3$ inversion lines.  The detection of one or more of these high density tracers attest to the presence of gas  with volume densities in excess of $10^{4}$~\cc.  The gas in this set of dust sources responsible for high excitation line emission is likely distributed within high density fragments that are unresolved by the Bolocam 33\arcsec\ angular resolution. Such inhomogeneous conditions are present in the sample of dust sources analyzed by \citet{beuther02} in which the mean volume densities range are $\sim$10$^{4-5}$~\cc\ but gas with volume densities of 10$^{5-6}$~\cc\ are inferred from multi-level transitions of CS. High angular resolution of one of these sources (IRAS 05358+3543) identifies at least 4 sub-fragments within the core with masses ranging from 1-18 \msun \citep{beuther07}.  It is reasonable to assume our sample of dust cores follow a similar clumpy distribution.  Therefore, the dust masses derived from the BGPS 1.1~mm dust emission are upper limits to the amount of high density material within the cloud.

To further isolate regions of high density within the cloud that are more coupled to the production of stars,  we examined the BGPS surface brightness images of 1.1~mm dust continuum emission and the images of source masks that link pixels to a cataloged source \citep{rosolowsky10}.  Many of the BGPS sources are marginally or well resolved by the 33\arcsec\ resolution. For each resolved source, pixels with surface brightness values in excess of 0.22~Jy/sr are consolidated.  This threshold corresponds to a mass surface density of 200~\mpcsq\ for a dust temperature of 14~K that approximately matches the K-band extinction threshold used by \citet{lada12}.  The set of  pixels that satisfy this limit are used to derive the solid angle, $\Omega_{\rm{dust,200}}$, mass, $M_{\rm{dust,200}}$, surface density, $\Sigma_{\rm{dust,200}}$, and mean volume density, $<n_{\rm{dust,200}}>$ of high column density material for each source.  The distributions of surface density, $\Sigma_{\rm{dust,200}}$, and volume density, $<n_{\rm{dust,200}}>$ are shown in Figure~\ref{fig1} and Figure~\ref{fig2}, respectively.  These segments within the dust cores have mean volume densities 10 times larger than the ambient cloud density values and are more representative of material linked to star formation than the composite BGPS sources. 

The distributions of GMC mass, $M_{\rm{GMC}}$, dust mass, $M_{\rm{dust}}$, and high column density dust mass, $M_{\rm{dust,200}}$ are shown in Figure~\ref{fig3}.  The absence of very massive (M$>$10$^{5.5}$~\msun) GMCs may reflect that in some cases, \texttt{CPROPS} identifies a massive segment of a larger molecular cloud.  Multiple large segments are common features in local giant molecular clouds.  For example,  the Orion molecular cloud complex is comprised of two major segments -- the Orion A and Orion B clouds.  At a distance of 15~kpc, the \coa\ $J=1-0$ emission from  these segments would be separately identified by the \texttt{CPROPS} algorithm at the sensitivities of the GRS.  The dust masses are typical of massive, dense fragments of clouds that are associated with massive star and stellar clusters formation \citep{battersby10}.

\begin{figure*}[htbp]
\includegraphics[width=6.5in]{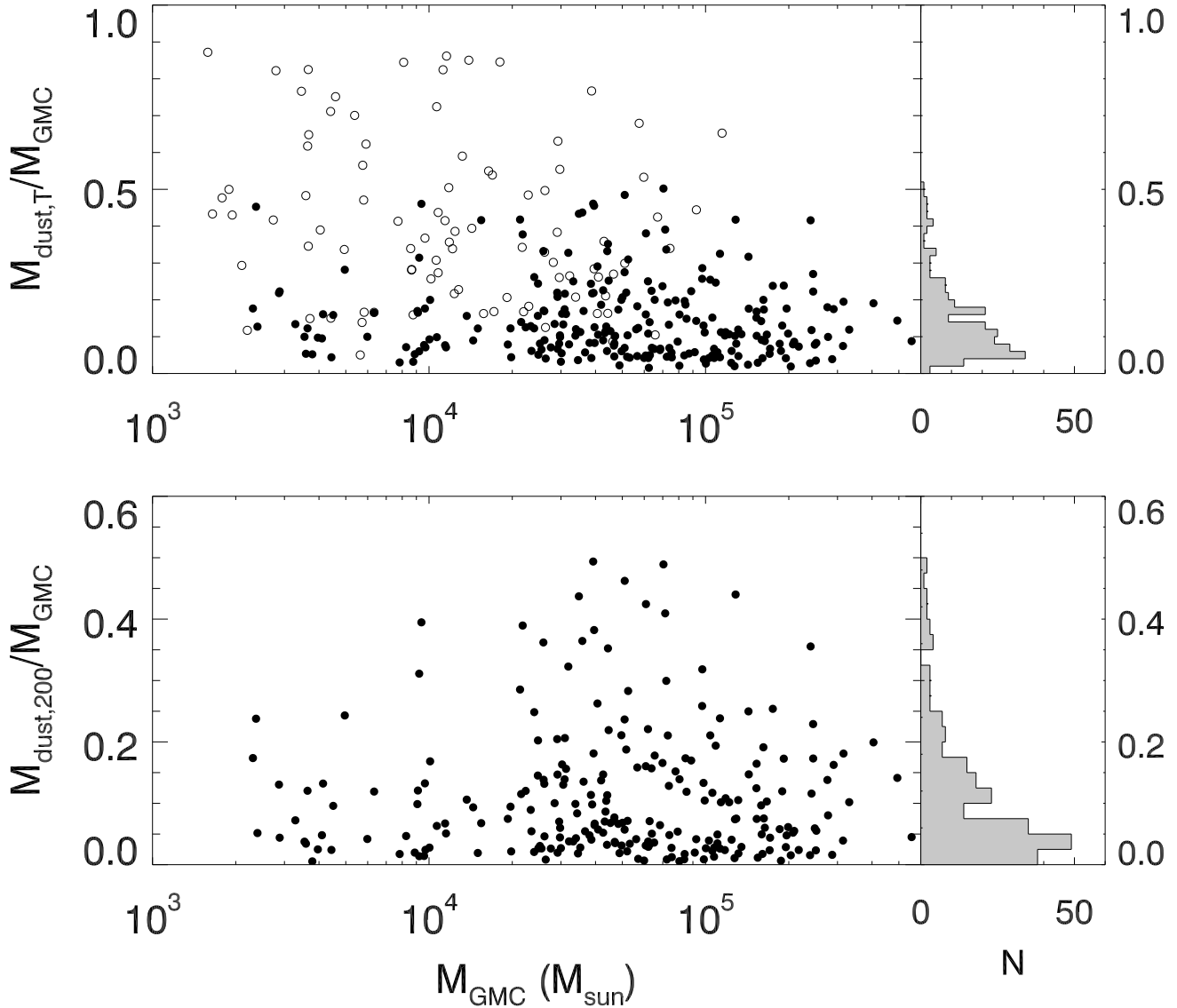}
\caption{
(\textit{Top}) The fraction of the GMC mass residing within BGPS dust sources, $M_{\rm{dust,T}}/M_{\rm{GMC}}$, as a function of GMC mass.  The solid and open circles correspond to clouds with angular radius $>$ 3\arcmin\ and $<$ 3\arcmin, respectively.  For clarity, individual error bars are not shown. The right hand panel show the distribution of $M_{\rm{dust,T}}/M_{\rm{GMC}}$ for clouds with angular size greater than 3'. (\textit{Bottom})  The dense gas fraction, \dgf\ as a function of GMC mass. The right hand panel show the distribution of \dgf\ values for clouds with angular size greater than 3'.  
}
\label{fig4}
\end{figure*}
\begin{figure*}[htbp]
\includegraphics[width=6.5in]{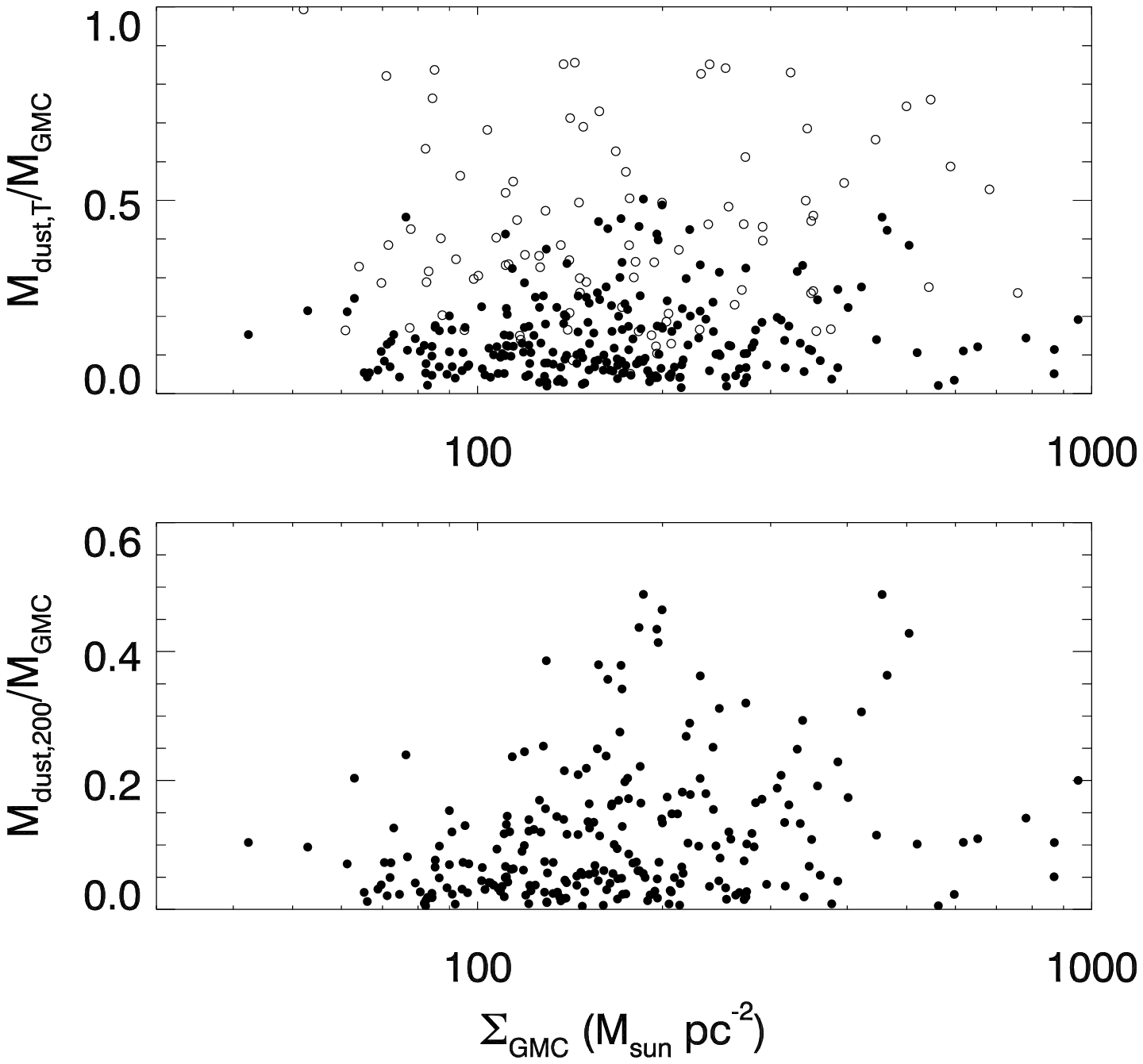}
\caption{
(\textit{Top}) The fraction of the GMC mass residing within BGPS dust sources, $M_{\rm{dust,T}}/M_{\rm{GMC}}$, as a function of GMC surface density. (\textit{Bottom}) The dense gas fraction, \dgf\ as a function of GMC mass surface density.}
\label{fig5}
\end{figure*}

\section{Dense Gas Fraction in GMCs}
The production of stars occurs within the localized, overdense regions within molecular clouds.  Therefore, the fractional mass contribution of the dense gas component sets an upper limit to the efficiency of star formation over the next free-fall time scale or longer.   In most cases, the dust sources are overdense regions with respect to the mean density of the GMC, as shown in Figure~\ref{fig2}, and are frequently associated with star formation activity \citep{dunham11a}.  \citet{battersby10} demonstrate that the dense, filamentary structures of clouds, exemplified by infrared dark clouds, are also well traced by the BGPS~1.1mm continuum emission.  The dense cores from which stars directly condense are embedded within these filaments. The fraction of the cloud mass that resides within the resident dust sources is $M_{\rm{dust,T}}/M_{\rm{GMC}}$ where $M_{\rm{dust,T}}$, is the total mass derived from of all primary and secondary BGPS sources within a cloud.  We estimate the dense gas mass fraction of a cloud, \dgf, as the ratio of mass residing within the high column density lines of sight and the GMC mass, $M_{\rm{dust,200}}/M_{\rm{GMC}}$.  As discussed in \S2.7 and \S3.1, the mass derived from the dust emission is an upper limit to the amount of material with densities greater than 10$^4$~\cc\ given the clumpy distribution of the dust sources.  The GMC masses derived from the \coa\ column density are likely lower limits due to our assumptions of excitation temperature, LTE, optical depth, and \coa\ abundance.  Therefore, our estimates of $M_{\rm{dust,T}}/M_{\rm{GMC}}$  and $M_{\rm{dust,200}}/M_{\rm{GMC}}$ are upper limits to the true mass fractions. It is worth emphasizing that these mass ratios are unaffected by uncertainties associated with our distance determination, since the distance terms cancel out of the ratio. Therefore, even if a cloud is misidentified as being on the near or far side of the Galaxy, this ratio will be unchanged.

The calculated fractions $M_{\rm{dust,T}}/M_{\rm{GMC}}$ and $M_{\rm{dust,200}}/M_{\rm{GMC}}$ as functions of cloud mass and cloud surface density are shown in Figure \ref{fig4} and Figure~\ref{fig5}, respectively. The values of $M_{\rm{dust,T}}/M_{\rm{GMC}}$ are affected by the BGPS data processing, which removes large scale structure to account for atmospheric contributions.  For clouds with angular sizes smaller than the angular extent of the Bolocam field of view of 6\arcmin\ (open circles), the median filtering removes the atmospheric contribution and recovers all or some large fraction of the cloud column density \citep{aguirre11, ginsburg13}.  In these cases, the derived dust masses are comparable to the \coa-derived cloud masses.  The median filtering is responsible for the results of \citet{dunham11b} who found decreasing volume densities with distance, which in most cases, corresponds to the decreasing angular size of the clouds.  For GMCs with sizes greater than 6\arcmin\ (filled points), this effective median filtering removes atmospheric and the extended cloud emission and recovers compact regions within the cloud with enhanced column density. The distributions of $M_{\rm{dust,T}}/M_{\rm{GMC}}$ and  $M_{\rm{dust,200}}/M_{\rm{GMC}}$ are asymmetrically skewed towards higher values.  The median value of $M_{\rm{dust,T}}/M_{\rm{GMC}}$ is $0.11_{-0.06}^{+0.12}$ where the errors correspond to 1$\sigma$ offsets that encompass 68\% of the sample.  Values of $M_{\rm{dust,200}}/M_{\rm{GMC}}$ are our best estimate for the dense gas mass fraction, \dgf, in GMCs.  The median value and 1$\sigma$ spread of the measured $M_{\rm{dust,200}}/M_{\rm{GMC}}$ distribution for GMCs with angular sizes greater than 3\arcmin\ are $0.07_{-0.05}^{+0.13}$. The measurement errors of these ratios are 0.06 and 0.04 for  $M_{\rm{dust,T}}/M_{\rm{GMC}}$ and $M_{\rm{dust,200}}/M_{\rm{GMC}}$ respectively with the largest contribution from the dust mass uncertainties. We re-emphasize that the derived ratios are limited by systematic errors that overestimate the true values of these mass fractions.

Recent studies by \citet{eden12, eden13} have analyzed the same data sets to derive $M_{\rm{dust,T}}/M_{\rm{GMC}}$ in a sample of clouds within varying environments in the Galactic Plane.  They found no systematic variation of mean $M_{\rm{dust,T}}/M_{\rm{GMC}}$ values between near side, far side, and tangent point clouds, although values range from 1-20\%.  \citet{eden13} considered the variation of this ratio between arm and inter-arm regions along the lines of sight over the range $37.83^\circ \le l \le 42.5^\circ$. They found no significant difference in clouds that lie within a spiral feature and those located between arms.  However, they did identify a very large value (36\%) for the Perseus arm, which is located on the far-side of the tangent point for these lines of sight.  This large value is most likely due to the small angular size of clouds at these distances and the effective median filtering of the BGPS processing that recovers  the cloud column density and mass rather than sub-regions of high volume or column density.  Excluding the distant, Perseus arm sample, their mean $M_{\rm{dust,T}}/M_{\rm{GMC}}$ values are 5\%. The factor of two difference between our results may lie in our application of a flux limit which biases our sample to larger dust masses or in the different cloud identification methods. 

Our results for $M_{\rm{dust},200}/M_{\rm{GMC}}$ can be further compared to other studies which estimate the dense gas fraction using infrared-derived extinctions of background stars as measures of gas column density.  When possible, such extinction studies are more reliable to evaluate a dense gas fraction than the method used in this study as the full range of cloud column density is derived with a singular tracer of gas column density and is not dependent on local, molecular chemistry or dust temperature. However, such extinction studies are challenging for more distance clouds owing to contamination by foreground stars as well as foreground extinctions.  To date, such studies are limited to a small number of local clouds and even a smaller number of GMCs in the molecular ring of the Galaxy, (although, see \citet{kainulainen13a}). From their sample of 11 local clouds, \citet{lada12} estimate the mass contained in regions with K-band dust extinctions greater than 0.8 magnitudes relative to the mass residing with $A_K > 0.1$ to be 0.10$\pm$0.06. \citet{kainulainen13} define a continuous dense gas fraction, $dM'(>N)= M(>N)/M_{{\rm TOT}}$ normalized to unity at a column density of 3$\times$10$^{21}$~\cmsq.  This function provides a concise description of the mass distribution in a cloud. Evaluated at 200 \mpcsq, they find low values ($\sim$0.02) in starless clouds and higher values 0.05-0.2 in star forming clouds like Cep~A, Perseus,  Taurus, and Orion.

\begin{figure*}[htbp]
\begin{center}$
\begin{array}{lll}
\includegraphics[width=2.2in]{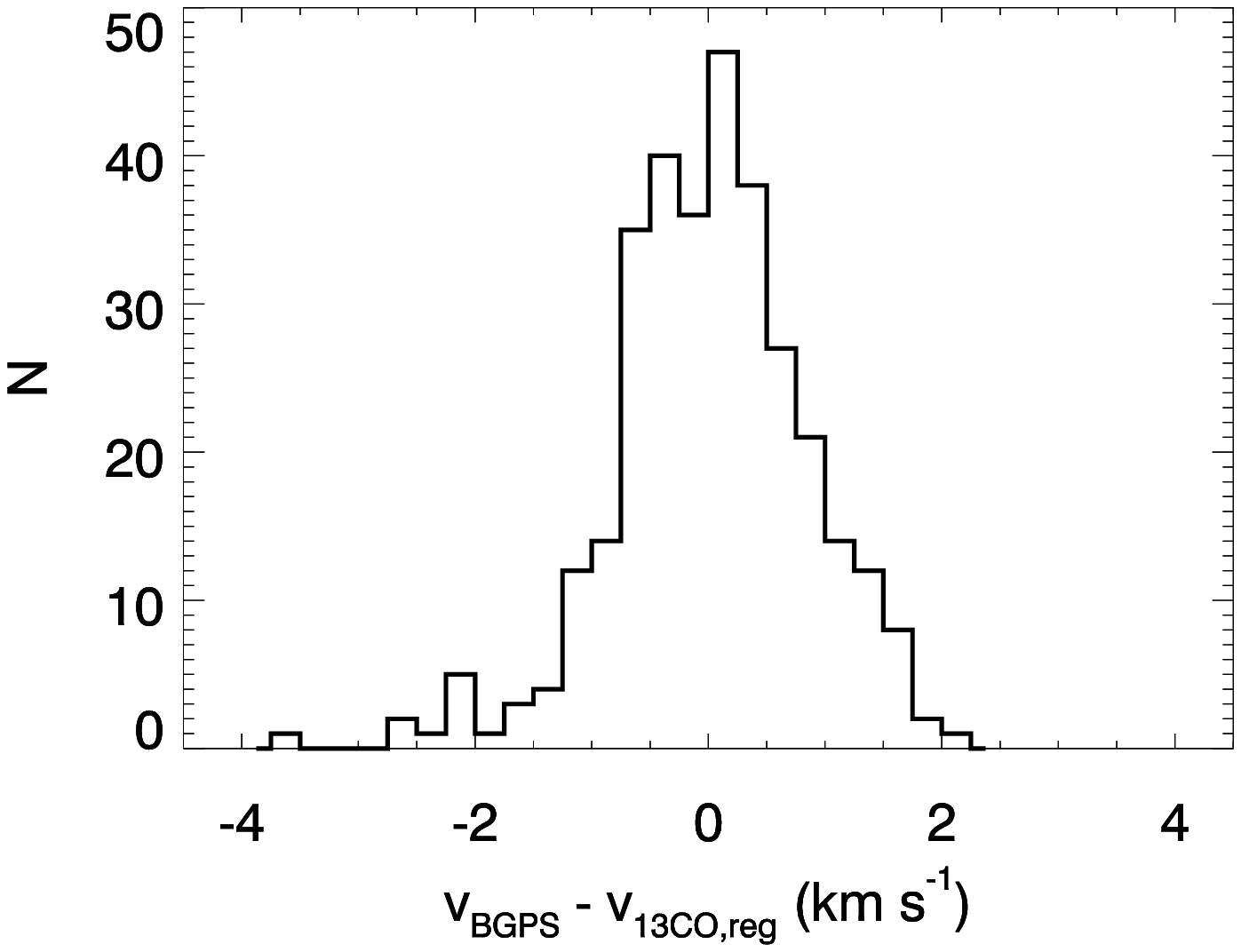} &
\includegraphics[width=2.2in]{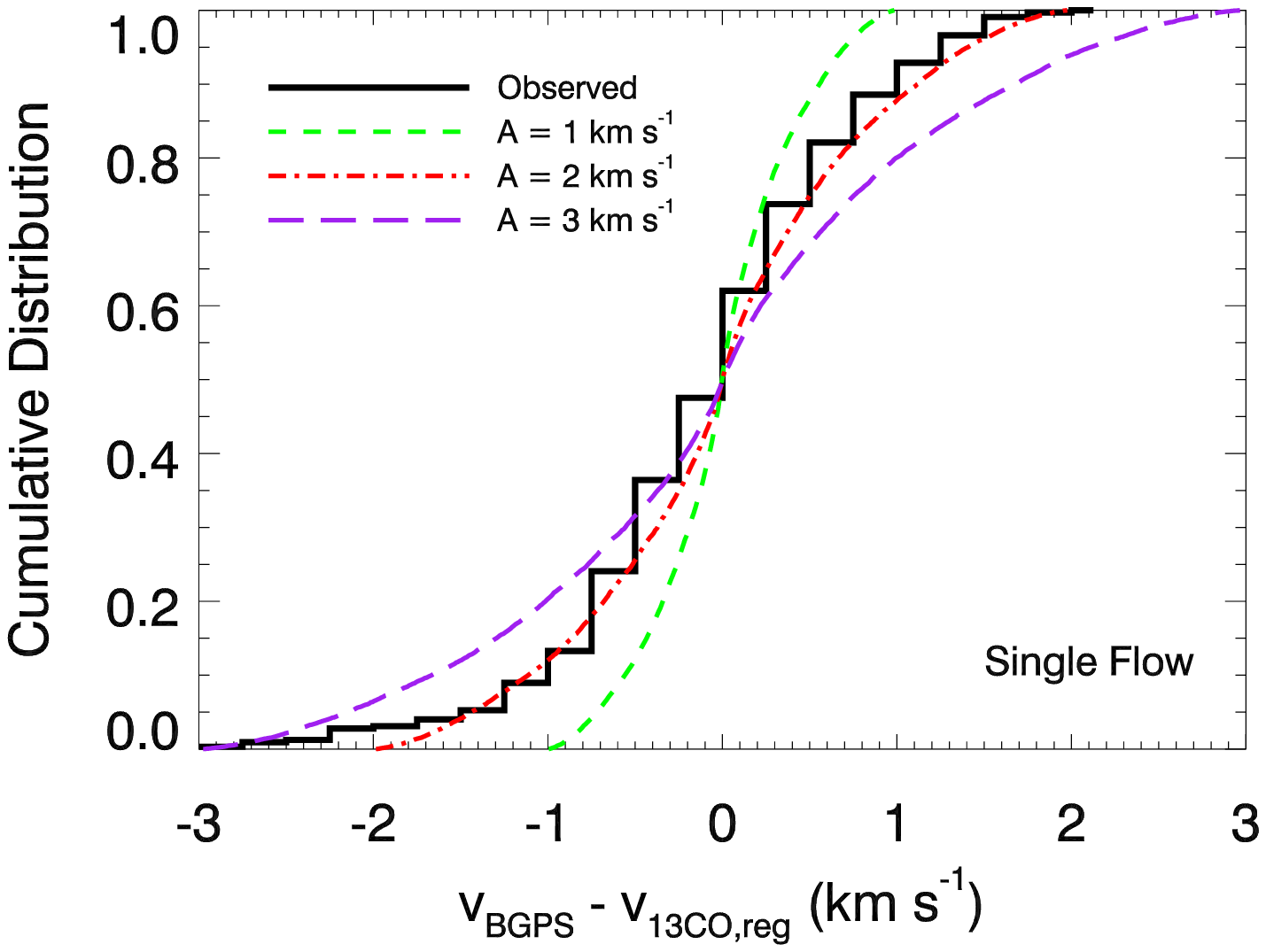} &
\includegraphics[width=2.2in]{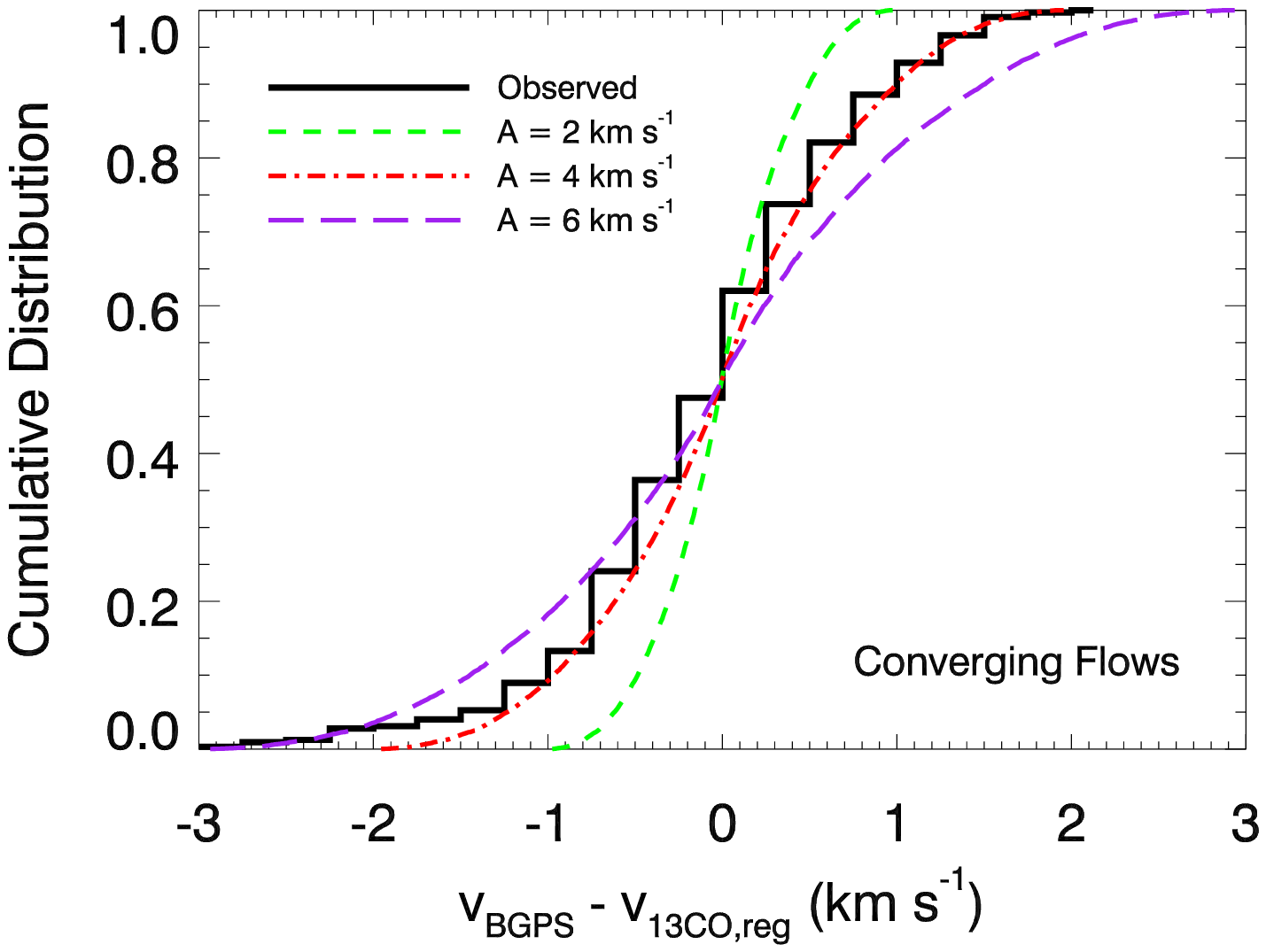} 
\end{array}$
\end{center}
\caption{a) Distribution of centroid velocities differences between the BGPS source and the \coa\ envelope. b) Model cumulative distribution of the line of sight velocity component of one-sided, infalling radial flows with random orientations to for amplitudes 1 (short dashed), 2 (dashed-dotted), and 3 (long dash) \kms.  The observed cumulative distribution is shown as the solid histogram.  c) Model cumulative distribution of the line of sight velocity component of converging, misaligned, radial flows with random orientations to for amplitudes 2 (short dashed), 4 (dashed-dotted), and 6 (long dashed) \kms. The models constrain the amplitudes of infalling flows to 2-4 \kms. }
 \label{fig6}
\end{figure*}

While our study does not quantify the star formation activity within this set of clouds, the derived values of \dgf\ have important implications to the transformation of gas into stars.  First, Figures \ref{fig4} and \ref{fig5} demonstrate that \dgf\ is independent of the mass and surface density of the host GMC.  This invariance excludes the suggestion by \citet{heiderman10}  that the extragalactic, super-linear Kennicutt-Schmidt scaling relationship emerges from a dependence of \dgf\ on $\Sigma_{\rm{mol}}=(A_{\rm{mol}}/A_{\rm{ap}})<\Sigma_{\rm{GMC}}>$, where $A_{\rm{mol}}/A_{\rm{ap}}$ is the area filling factor of GMCs within the extragalactic aperture and $<\Sigma_{\rm{GMC}}>$ is the average mass surface density of GMCs. Furthermore, the mean value of the dense gas mass fraction can be compared to the fraction of cloud material converted into stars over a free-fall time, $\epsilon_{FF}$, which provides an important constraint to theoretical descriptions of star formation. For a volumetric star formation law, \citet{krumholz12} calculate the range  $0.003 < \epsilon_{FF} < 0.03$ that describes the composite measurements of the Kennicutt-Schmidt scaling relationship from resolved GMCs in the Galaxy and nearby galaxies to more distant starbursts and high redshift systems. Figures~\ref{fig4} and \ref{fig5} demonstrate a small amount of a cloud's mass resides within a dense configuration to form newborn stars.  Recalling that these are likely upper limits and only a fraction of the dense gas will develop into stars, the small values of $\epsilon_{FF}$ can be attributed to the deficiency of high volume density gas as measured by \dgf, rather than some process {\it within} the dense gas that inhibits infall or the presence of a diffuse, molecular component that is inert to the production of stars.  Star formation is therefore limited by the processes responsible for the formation of dense, compact structures within magneto-turbulent clouds.

\section{Kinematics of Dense Regions}
The kinematic relationship between the dense gas and the low density substrate, as traced by \coa\ emission, offers insight to processes responsible for clump and filament formation in GMCs. Many recent studies have proposed that dense filamentary structures develop at the stagnation points of supersonic, converging  or collapsing gas streams in GMCs \citep{padoan2002, krumholz05, ballesteros2011, gong09, gong11}.  In such cases, there should be a velocity offset between the dense gas and the converging, low density material \citep{smith12}.  To evaluate the velocity displacement between the BGPS source and the surrounding, low density material, we calculate the difference between the centroid velocity of the dense gas tracer from \citet{schlingman11} or \citet{dunham11b} and the centroid velocity of the \coa\ spectrum averaged over a 3$\times$3 arcmin$^2$ area centered on the BGPS source.   We restrict the sample to primary BGPS sources residing within GMCs with angular extents larger than 3\arcmin\ so that the \coa\ velocity centroid is representative of local motions about the BGPS source rather than the entire cloud. The distribution of velocity centroid differences for our sample is shown in Figure~\ref{fig6}a. It is strongly peaked on zero velocity displacement.  Both \citet{walsh04} and \citet{kirk10} find similar distributions of velocity differences between the velocity centroids of \coa\ and high density tracers in the nearby Perseus molecular cloud. \citet{wienen2012} also derive a centrally peaked distribution of velocity differences centered on zero between \coa\ and NH$_3$ line centroids for a flux limited sample of ATLASGAL sources throughout the inner Galaxy. 

The small velocity centroid differences found in these distributions do not necessarily indicate small velocity displacements between the dense gas and overlying material or that these two components are co-moving.  Each velocity centroid measures the line of sight component of the respective gas motions.  Since it is reasonable to assume that any relative motion between the dense gas and low density medium is randomly oriented with respect to the observer, true velocity 
displacements are necessarily diluted by orientation. 

We model the observed distribution  of velocity differences in the context of converging, radial flows with varying velocity offsets that are randomly oriented. While such converging flows are not a unique description of the velocity differences between the dense gas and lower density medium, these models can constrain the mean flow velocity in the limiting case that the \coa\ $J=1-0$ emission primarily originates within a converging or collapsing parcel of low density gas centered on the BGPS source. Two simple flow configurations are considered: a single, one sided flow, and a two-sided, converging, yet misaligned flow.  The flow velocity is referenced to the velocity of the dense gas and is taken as the average projected velocity to the observed line of sight. In practice, the mean projected velocity also depends on the density of the flows being considered as this impacts the line excitation. 

For a one-sided, radial flow of low density gas moving relative to the dense gas with a random inclination angle, the  radial velocity is 
\begin{equation} \label{single_flow}
v_r = A \cos\theta \sin\phi\ ,
\end{equation}
where $A$ is the velocity amplitude of the flow, $\theta$ is the angle of the flow off the x-y plane and $\phi$ is the polar angle within the x-y plane. The observer's line of sight is along the x axis. The value of $\cos\theta$ is selected randomly from a uniform distribution range $-1<\cos\theta<1$. The value of $\phi$ is selected randomly from a uniform distribution between $0<\phi<2\pi$.  The cumulative distributions of relative velocities for 3 different values of $A$ are shown in Figure \ref{fig6}b along with the observed cumulative distribution. The model distributions are centered on zero with widths dependent on the value of $A$. The observed distribution is well bounded by flow velocities of 1-3~\kms.  Models with $A=1$ and 3~\kms\ respectively under-predict and over-predict the frequency of large velocity centroid differences. The model with $A=2$~\kms\ offers a reasonable approximation of the measured distribution. 

The second set of models considers radially converging flows that are misaligned with respect to each other. Such misaligned flows are present in the computational simulations of supersonic turbulent flows in GMCs  \citep{padoan2002,smith12}. The projected velocity is determined from equation \eqref{single_flow} for the front, (F), and back,(B), hemispheres such that  $0<\cos\theta_F<1$ and $-1<\cos\theta_B<0$ respectively. The model projected velocity of the low density gas is the average of the radial velocity of the two flows, $(v_{r,F}+v_{r,B})/2$. The cumulative distributions of relative velocities for three different velocity amplitudes are shown in Figure \ref{fig6}c along with the observed distribution. The distribution of relative velocities in this case is narrower than the one-sided flow case for the same value of $A$.   When the two-sided, radial flows are nearly aligned, the velocity difference is always close to zero as the receding frontside and approaching backside velocities nearly cancel.  For the two-sided, misaligned flow, the velocity amplitude of $A=4$~\kms\ best matches the observed distribution of velocity differences.  

In the context that the measured distribution of velocity displacements results from converging or infalling material towards a dense, central clump or core, these simple models constrain the velocity of the converging gas to be 2-4~\kms. We emphasize that this is a statistical measure of the velocity difference averaged over random orientations to the line of sight. It is not possible to derive a true velocity displacement between the dense gas and low density material circumscribing the dense gas for any single target.

The converging gas motions inferred from the models imply a Mach number of 10-20 for a gas temperature of 10~K. For an for isothermal gas, such  converging, supersonic motions should produce very large density fluctuations in shocks.  However, on the spatial scales resolved by these observations, such shocks would be difficult to detect.  Indeed, we measure no systematic variation of the dense gas fraction or a contrast ratio, ($\Sigma_{\rm{dust},200}/\Sigma_{\rm{GMC}}$ or $n_{\rm{dust},200}/n_{\rm{GMC}}$), with velocity dispersion. 

Recent computational studies have examined the role of converging flows in the 
formation of molecular clouds and the development of cloud sub-structure and 
turbulence \citep{heitsch08,hennebelle08,banerjee09, gong09, gong11, clark12}. A common feature in the simulations is the development of massive filaments as material accumulates within the post-shock regions due to gravity, turbulence, and thermal instabilities.  The filaments are gravitationally unstable and undergo 
further collapse and fragmentation into localized cores. \citet{gong11} carried out a parameter study of converging, supersonic flows for Mach numbers ranging from 1 to 9. They find higher Mach number flows generate a larger range of core masses and a smaller median core mass due to the reduced Jeans' mass in the post shock region. Our estimate of the converging flow Mach number is comparable to their largest Mach number used in their parameter study. Nevertheless, the maximum mass from their simulation is 50~\msun, which is much smaller than the typical dust mass listed in Table 1. 

In the case of converging streams of warm, neutral, atomic gas (WNM) studied by \citet{heitsch08}, \citet{hennebelle08}, \citet{banerjee09}, and \citet{clark12} the formation of filaments is a downstream process resulting from instabilities and turbulence generated from the large scale collision and occurring 1-2$\times10^7$ years after the initial interaction. The clumps and filaments have Mach numbers and surface densities comparable to the BGPS sources studied here but with maximum masses of $\sim500$~\msun, which are much smaller than the typical mass of a BGPS source \citep{banerjee09}.  More massive filaments may result from the continued accumulation of material from the large scale collapse of the clouds predicted by these simulations. Unfortunately, a direct comparisons between the gas properties and kinematics associated with these regions to our results is not possible or straightforward as the simulations focus on mass distributions of the densest regions (cores) in the simulations, which we do not resolve, and the true velocities of large scale flows, which we cannot determine. Despite this, the results of \citet{clark12} suggest low fractions of dense gas in colliding flows, consistent with our findings. At the onset of star formation in their simulations they find $M(n>10^3$~\cc)$/M(n>100$~\cc) to be $\sim1-2\%$ for slow colliding flows, with $v_{\rm{flow}}=6.8$~\kms, and $\sim20-25\%$ for fast colliding flows, with $v_{\rm{flow}}=13.6$~\kms.

\section{Conclusions}

We have examined the properties of the host giant molecular clouds associated with a flux limited sample of 1.1~mm dust continuum sources from the Bolocam Galactic Plane Survey.  These dust sources are representative of massive clumps in GMCs from which stellar clusters and massive stars ultimately form. The fraction of the GMC mass residing within the dust continuum sources is $0.11_{-0.06}^{+0.12}$.  The fraction of GMC mass residing within the dust sources with mass surface densities greater than 200 \mpcsq\ is $0.07_{-0.05}^{+0.13}$.  This dense gas fraction is independent of GMC mass and suggest that the inefficiency of star formation is caused by the deficiency of dense gas. The distribution of velocity differences between the dense gas and surrounding lower density material is centrally peaked on zero displacement.  Models of radially converging flows with random orientations to the observer constrains the velocity of these flows to 2-4 \kms. 

We acknowledge support from NSF grant AST-1009049. The authors thank Ashley Bemis for preliminary work on this project, Jouni Kainulainen and Ron Snell for helpful discussions, and the referee whose suggestions improved the content of this work.  This publication makes use of molecular line data from the Boston University-FCRAO Galactic Ring Survey (GRS). The GRS is a joint project of Boston University and Five College Radio Astronomy Observatory, funded by the National Science Foundation under grants AST-9800334, AST-0098562, AST-0100793, AST-0228993, and  AST-0507657.  The BGPS project was supported in part by NSF grant AST-0708403, and was performed at the Caltech Submillimeter Observatory (CSO), supported by NSF grants AST-0540882 and AST-0838261. This research has made use of the NASA/ IPAC Infrared Science Archive, which is operated by the Jet Propulsion Laboratory, California Institute of Technology, under contract with the National Aeronautics and Space Administration.

\clearpage
\LongTables
\begin{deluxetable*}{cccrrrrrrrrc}
\tablecolumns{11} 
\tablewidth{555pt}
\tabletypesize{\scriptsize}
\tablecaption{GMC Properties and Dust Masses} 
\label{table1}
\tablehead{
 \colhead{Primary} &
 \colhead{$T_{\rm{thr}}$} &
 \colhead{$l$} &
 \colhead{$b$} &
 \colhead{$V_{\rm{LSR}}$} &
 \colhead{\dkin} &
 \colhead{$R$} &
 \colhead{$\sigma_v$}&
 \colhead{$M_{\rm{GMC}}$}&
 \colhead{$M_{\rm{dust,T}}$}&
 \colhead{$M_{\rm{dust},200}$} &
 \colhead{\# BGPS}\\
 \colhead{{BGPS}} &
 \colhead{{(K)}} &
 \colhead{{(deg)}} &
 \colhead{{(deg)}} &
 \colhead{{(km/s)}}&
 \colhead{{(kpc)}}&
 \colhead{{(pc)}} &
 \colhead{{(km/s)}}&
 \colhead{{($10^3 M_\odot$)}} &
  \colhead{{($10^3M_\odot$)}} &
  \colhead{{($10^3M_\odot$)}} & 
 \colhead{{in GMC}}
}
\startdata
2430&  2.6& 18.621& -0.093&  45.5& 3.64&  8.86 (  0.14)& 1.36 ( 0.01)& 43.5 (  0.7)&  5.0 (  1.0)&  1.9 (  0.8)&  6\\
2431&  1.0& 18.668&  0.035&  80.0&10.65& 11.82 (  0.02)& 1.44 ( 0.01)& 56.6 (  0.1)& 10.3 (  4.3)& 10.3 (  4.3)&  1\\
2436&  3.6& 18.706& -0.231&  43.5&12.54& 16.96 (  2.54)& 2.94 ( 0.19)&290.5 (72.0)& 50.8 (15.2)&  8.2 (  3.4)&  3\\
2439&  1.0& 18.693&  0.012&  27.2&13.67& 16.60 (  2.06)& 1.64 ( 0.18)& 77.9 (17.1)& 15.5 (  6.5)& 15.5 (  6.5)&  1\\
2446&  0.7& 18.683&  0.287&  19.5&14.29& 33.54 (  3.78)& 2.43 ( 0.06)&555.9 (63.7)& 48.9 (11.2)& 20.4 (  9.0)&  5\\
2453&  1.0& 18.820& -0.076& 122.6& 7.62&  8.91 (  4.47)& 1.61 ( 0.64)& 17.2 (14.8)&  1.8 (  0.8)&  1.8 (  0.8)&  1\\
2456&  1.2& 18.841& -0.295&  42.7&12.64&  5.79 (  1.13)& 1.01 ( 0.06)&  8.9 (  1.3)&  9.1 (  4.0)&  9.1 (  4.0)&  1\\
2460&  2.5& 18.919& -0.356&  65.0& 4.45& 25.11 (  0.34)& 2.30 ( 0.06)&330.7 (24.4)& 39.4 (  5.6)&  1.6 (  0.7)& 22\\
2464&  0.8& 18.920& -0.401&  25.5& 2.23&  3.26 (  0.21)& 0.89 ( 0.05)&  3.5 (  0.2)&  0.4 (  0.1)&  0.2 (  0.1)&  2\\
2469&  2.9& 18.887&  0.041&  49.2&12.25& 12.71 (  0.22)& 1.37 ( 0.32)&131.2 (16.5)& 15.7 (  6.6)& 15.7 (  6.6)&  1
\enddata
\tablecomments{Columns list the (1) ID of the primary source in the BGPS catalog, (2) threshold temperature used to define the GMC, (3) Galactic longitude (4) Galactic latitude, (5) local standard of rest (LSR) velocity of the GMC, (6) kinematic distance of the GMC, (7) radius of the GMC, (8) velocity dispersion of the GMC, (9) GMC mass, (10) total mass of BGPS sources, (11) total mass of BGPS source pixels with $\Sigma_{\rm{dust}}>200$~\mpcsq, and (12) total number of BGPS sources within the GMC. Numbers in parentheses indicate uncertainties of the listed quantities.
\newline (The full table will be available in its entirety in a machine-readable form in the online journal. A portion is shown here for guidance regarding its form and content.)}
\end{deluxetable*} 
 \clearpage

\begin{thebibliography}{}
\bibitem[Aguirre et al.(2011)]{aguirre11} Aguirre, J. E., Ginsburg, A. G., Dunham, M. K., et al. 2011, ApJS, 192, 4
\bibitem[Ballesteros et al.(2011)]{ballesteros2011} Ballesteros-Paredes, J. Hartmann, L.~W., Vazquez-Semadeni, E., Heitsch, F., \& Zamora-Aviles, M.~A. 2011, \mnras, 411, 65
\bibitem[Banerjee et al.(2009)]{banerjee09} Banerjee, R., Vazquez-Semadeni, E., Hennebelle, P., \& Klessen, R.~S. 2009, \mnras, 398, 1082
\bibitem[Battersby et al.(2010)]{battersby10} Battersby, C., Bally, J., Jackson, J. M., et al. 2010, \apj, 721, 222
\bibitem[Beuther et al.(2002)]{beuther02} Beuther, H., Schilke, P., Menten, K. M., Motte, F., Sridharan, T. K., \& Wyrowski, F. 2002, \apj, 566, 945
\bibitem[Beuther et al.(2007)]{beuther07} Beuther, H., Leurini, S., Schilke, P., Wyrowski, F., Menten, K. M., \& Zhang, Q. 2007, A\&A, 466, 1065
\bibitem[Bigiel et al.(2008)]{bigiel08} Bigiel, F., Leroy, A., Walter, F., et al. 2008, AJ, 136, 2846
\bibitem[Blake et al.(1987)]{blake1987} Blake, G. A., et al. 1987, \apj, 315, 621
\bibitem[Brunt et al.(2003)]{brunt03} Brunt, C. M. 2003, \apj, 584, 293
\bibitem[Clark et al.(2012)]{clark12} Clark, P. C., Glover, S. C., Klessen, R. S., \& Bonnell, I. 2012, MRAS, 424, 2599
\bibitem[Clemens(1985)]{clemens85} Clemens, D. P. 1985, \apj, 295, 422
\bibitem[Dickman(1978)]{dickman78} Dickman R. L., 1978, ApJS, 37, 407
\bibitem[Dunham et al.(2011a)]{dunham11a} Dunham, M. K., Robitaille, T. P., Evans, N. J., II, et al. 2011a, \apj, 731, 90
\bibitem[Dunham et al.(2011b)]{dunham11b} Dunham, M. K., Rosolowsky, E., Evans, N. J., II, et al. 2011b, \apj, 741, 110
\bibitem[Eden et al.(2012)]{eden12} Eden D. J., Moore T. J. T., Plume R., \& Morgan L. K. 2012, MNRAS, 422, 3178
\bibitem[Eden et al.(2013)]{eden13} Eden D. J., Moore T. J. T., Morgan L. K.,  Thompson, M. A., \& Urquhart, J. S. 2013, MNRAS, 431, 1587
\bibitem[Ellsworth-Bowers et al. (2013)]{ellsworth2013} Ellsworth-Bowers, T.P., Glenn, J., Rosolowsky, E., Mairs, S., Evans, N. J., II, Battersby, C., Ginsburg, A., Shirley, Y.~L., Bally, J. 2013, \apj, 770, 39
\bibitem[Gao \& Solomon(2004)]{gao04} Gao, Y., \& Solomon, P. M. 2004, \apj, 606, 271
\bibitem[Ginsburg et al.(2013)]{ginsburg13} Ginsburg, A., Glenn, J., Rosolowsky, E., et al. 2013, APJS, 208, 14
\bibitem[Goldsmith et al.(2007)]{goldsmith07} Goldsmith, P. F., Li, D., \& Krco, M. 2007, \apj, 654, 273
\bibitem[Gong \& Ostriker(2009)]{gong09} Gong, H. \& Ostriker, E.~C. 2009, \apj, 699, 230 
\bibitem[Gong \& Ostriker(2011)]{gong11} Gong, H. \& Ostriker, E.~C. 2011, \apj, 729, 120 
\bibitem[Gutermuth et al.(2011)]{gutermuth11} Gutermuth, R. A., Pipher, J. L., Megeath, S. T., Myers, P. C., Allen, L. E., \& Allen, T. S. 2011. \apj, 739, 84
\bibitem[Heiderman et al.(2010)]{heiderman10} Heiderman, A., Evans, N. J., II, Allen, L. E., Huard, T., \& Heyer, M. H. 2010. \apj, 723, 1019
\bibitem[Heyer et al.(2004)]{heyer04} Heyer, M. H., Corbelli, E., Schneider, S. E., \& Young, J. S. 2004, \apj, 602, 723
\bibitem[Heyer \& Brunt(2004)]{heyer&brunt04} Heyer, M. H., \& Brunt, C. M. 2004, \apj, 615, L45
\bibitem[Heyer et al.(2009)]{heyer09} Heyer, M. H., Krawczyk, C., Duval, J., \& Jackson, J. M. 2009, \apj, 699, 1092
\bibitem[Heitsch et al.(2008)]{heitsch08} Heitsch, F., Hartmann, L.~W., Slyz, A.~D., Devriendt, J.~E.~G., \& Burkert, A. 2008, ApJ, 674, 316
\bibitem[Hennebelle et al.(2008)]{hennebelle08} Hennebelle, P., Banerjee, R., Vazquez-Semadeni, E., Klessen, R. S., \& Audit, E. 2008, A\&A, 486, L43
\bibitem[Jackson et al.(2002)]{jackson02} Jackson, J. M., Bania, T. M., Simon, R., et al. 2002, \apj, 566, L81
\bibitem[Jackson et al.(2006)]{jackson06} Jackson, J. M., et al. 2006, ApJS, 163, 145
\bibitem[Kainulainen \& Tan(2013)]{kainulainen13} Kainulainen, J., \& Tan, J. C. 2013, A\&A, 549, A53
\bibitem[Kainulainen et al.(2013a)]{kainulainen13a} Kainulainen, J., Ragan, S.~E., Henning, T., \& Stutz, A. 2013, A\&A, 557, A120
\bibitem[Kennicutt(1989)]{kennicutt89} Kennicutt, R. C., Jr. 1989, \apj, 344, 685
\bibitem[Kennicutt(1998)]{kennicutt98} Kennicutt, R. C., Jr. 1998, \apj, 498, 541
\bibitem[Kennicutt \& Evans(2012)]{kennicutt12} Kennicutt, R. C., Jr., \& Evans, N. J., II 2012, ARA\&A, 50, 531
\bibitem[Kirk et al.(2010)]{kirk10} Kirk, H., Pineda, J., Johnstone, D., \& Goodman, A. 2010, \apj, 723, 457
\bibitem[Krumholz \& McKee(2005)]{krumholz05} Krumholz, M. \& McKee, C.F. 2005, \apj, 630, 250
\bibitem[Krumholz et al.(2012)]{krumholz12} Krumholz, M. R., Dekel, A., \& McKee C. F. 2012, \apj, 745, 69
\bibitem[Lada et al.(2010)]{lada10} Lada, C. J., Lombardi, M., \& Alves J. F. 2010, \apj, 724, 687
\bibitem[Lada et al.(2012)]{lada12} Lada, C. J., Forbrich, J., Lombardi, M., \& Alves J. F. 2012, \apj, 745, 190
\bibitem[Larson(1981)]{larson81} Larson, R. B. 1981, \mnras, 194, 809
\bibitem[Leroy et al.(2013)]{leroy13} Leroy, A. K., Walter, F., Sandstrom, K., et al. 2013, AJ, 146, 19 
\bibitem[Milam et al.(2005)]{milam05} Milam, S. N., Savage, C., Brewster, M. A., Ziurys, L. M., \& Wyckoff, S. 2005, \apj, 634, 1126
\bibitem[Ossenkopf \& Henning(1994)]{ossenkopf94} Ossenkopf, V., \& Henning, T. 1994, A\&A, 291, 943
\bibitem[Padoan et al.(2000)]{padoan00} Padoan, P., Juvela, M., Bally, J., \& Nordlund, \ang. 2000, ApJ, 529, 259
\bibitem[Padoan \& Nordlund(2002)]{padoan2002} Padoan, P. \& Nordlund, \ang. 2002, \apj, 576, 870
\bibitem[Ripple et al.(2013)]{ripple13} Ripple, F., Heyer, M. H., Gutermuth, R., Snell, R. L., Brunt, C. M. 2013, MNRAS, 431, 1296
\bibitem[Roman-Duval et al.(2010)]{roman-duval2010} Roman-Duval, J., Jackson, J.~M., Heyer, M. Rathborne, J., 
\& Simon, R., 2010, \apj, 723, 492
\bibitem[Rosolowsky \& Leroy(2006)]{rosolowsky06} Rosolowsky, E., \& Leroy, A. 2006, PASP, 118, 590
\bibitem[Rosolowsky et al.(2010)]{rosolowsky10} Rosolowsky, E., Dunham, M. K., Ginsburg, A., et al. 2010, ApJS, 188, 123
\bibitem[Schlingman et al.(2011)]{schlingman11} Schlingman, W. M., Shirley, Y. L., Schenk, D. E., et al. 2011, ApJS, 195, 14
\bibitem[Shetty et al.(2009)]{shetty09} Shetty, R., Kauffmann, J., Schnee, S., Goodman, A. A., Ercolano, B., 2009, \apj, 696, 2234
\bibitem[Shetty, Kelly, \& Bigiel(2013)]{shetty13} Shetty, R., Kelly, B.~C., \& Bigiel, F. 2013 \mnras, 430, 288
\bibitem[Smith et al.(2012)]{smith12} Smith, R. J., Shetty, R., Stutz, A. M., \& Klessen, R. S. 2012, \apj, 750, 64
\bibitem[Solomon et al.(1987)]{solomon1987} Solomon, P.~M., Rivolo, A.R., Barret, J., \& Yahil, A. 1987, 
\apj, 319, 730
\bibitem[Stil et al.(2006)]{stil06} Stil, J. M., Taylor, A. R., Dickey, J. M., et al. 2006 AJ, 132, 1158 
\bibitem[van Dishoeck \& Black(1989)]{van89} van Dishoeck, E. F., \& Black, J. H. 1989, \apj, 340, 273
\bibitem[Visser et al.(2009)]{visser09} Visser, R., van Dishoeck, E. F., \& Black, J. H. 2009, A\&A, 503, 323
\bibitem[Walsh et al.(2004)]{walsh04} Walsh, A., Myers, P., \& Burton, M. 2004, \apj, 614, 194
\bibitem[Wienen et al.(2012)]{wienen2012} Wienen, M., Wyrowski, F., Schuller, F., Menten, K.M., Walmsley, C.M., Bronfman, L, \& Motte, F. 2012, A\&A, 544, 146
\bibitem[Wong, \& Blitz (2002)]{wong2002} Wong, T., \& Blitz, L. 2002, \apj, 569, 157
\bibitem[Wu et al.(2005)]{wu05} Wu, J., Evans, N. J., Gao, Y., et al. 2005, \apj, 635, L173
\end{thebibliography}
\end{document}